\documentstyle[twoside,fleqn,npb,amsfonts,epsfig]{article}
\newcommand{\bea}{\begin{eqnarray}}
\newcommand{\eea}{\end{eqnarray}}
\newcommand{\ba}{\begin{array}}
\newcommand{\ea}{\end{array}}
\newcommand{\nn}{\nonumber}
\newcommand{\lsim}{\raisebox{-0.13cm}{~\shortstack{$<$ \\[-0.07cm] $\sim$}}~}

\newcommand{\dis}{\displaystyle}

\renewcommand{\b}{{\mathbb B}}
\renewcommand{\c}{{\mathbb C}}
\newcommand{\cbar}{\bar{\mathbb C}}
\newcommand{\AmS}{{\protect\the\textfont2
  A\kern-.1667em\lower.5ex\hbox{M}\kern-.125emS}}

\title{Predictions for Lepton Flavour Violation in $Z$ decays\thanks{
    Talk given at Loops and Legs 2000, Bastei, Germany, April 9-14.}}
\author{
    \underline{Jos\'e I. Illana}$^{\rm \small a}$\thanks{
    On leave from Departamento de F{\'\i}sica Te\'orica y del Cosmos,
    Universidad de Granada, Spain.}
and T. Riemann\address{
Deutsches Elektronen-Synchrotron DESY, \\
Platanenallee 6, D-15738 Zeuthen, Germany
\vspace{-4cm}\\
 DESY 00--072\\
 hep-ph/0006055\\
 Version of 28.09.00
\vspace{3cm}
}}

\begin{document}

\begin{abstract}
Recent experimental results suggest that the neutrinos of the Standard
Model are massive, though light. Therefore they may mix with each other giving
rise to lepton flavour or even lepton number violating processes, depending
on whether they are Dirac or Majorana particles. Furthermore, 
the lightness of the observed neutrinos may be explained by the existence of
heavy ones, whose effects on LFV would be very sizeable. We present an
analysis of the effect of massive neutrinos on flavour-changing decays of the 
$Z$ boson into leptons, at the one-loop level, independent of neutrino mass 
models. Constraints from present experiments are taken into account. 
\end{abstract}

\maketitle

\section{Introduction and motivation}

With the Giga--$Z$ option of the Tesla linear collider project one may expect 
the production of about $10^{9}$ $Z$ bosons at resonance \cite{Hawkings:1999ac}.
This huge rate, about a factor 100 higher than at LEP~1, allows 
one to study a number of problems with unprecedented precision. 
Among them is the search for lepton-flavour changes in $Z$ decays:
$Z \to e\mu, \mu\tau, e\tau$. 
The best {\em direct} limits are obtained by searches at LEP~1 
(95\% c.l.) \cite{PDG:1998aa}: 
\bea
\label{lep-em}
{\rm BR}(Z\to e^{\mp}\mu^{\pm}) &<& 1.7 \times 10^{-6} 
,
\\
\label{lep-et}
{\rm BR}(Z\to e^{\mp}\tau^{\pm}) &<& 9.8 \times 10^{-6}
,
\\
\label{lep-tm}
{\rm BR}(Z\to \mu^{\mp}\tau^{\pm}) &<& 1.2 \times 10^{-5}
.
\eea
A careful analysis shows that the sensitivities could be largely improved 
at the Giga--$Z$ \cite{Wilson:1998bb},
\bea
\label{lep-emx}
{\rm BR}(Z\to e^{\mp}\mu^{\pm}) &<&  2 \times 10^{-9},
\\
\label{lep-etx}
{\rm BR}(Z\to e^{\mp}\tau^{\pm}) &<& f \times 6.5 \times 10^{-8},
\\
\label{lep-tmx}
{\rm BR}(Z\to \mu^{\mp}\tau^{\pm}) &<& f \times 2.2 \times 10^{-8},
\eea
with $f = 0.2 \div 1.0$.  

Non-zero rates are expected if neutrinos are massive and mix 
\cite{Pontecorvo:1957cp}.
From experiments we have evidence of tiny neutrino masses and substantial 
mixings. Unnaturally small mass scales may be indicative for a mechanism which 
produces at the same time very large masses. 
Heavy neutrinos are introduced in most GUTs \cite{Langacker:1981js} and
string-inspired models 
\cite{Hewett:1989xc}, and are suggested by the seesaw mechanism
\cite{seesaw}.

The above observations motivate us to have a closer look at
the prospects of observing LFV. We will explore here the following 
scenarios: 

(i) The $\nu$SM. We treat the known $n_G=3$ generations of {\em light 
neutrinos} ($\nu_e,\nu_\mu, \nu_\tau$) as {\em massive Dirac} particles.
Individual lepton numbers $L_e, L_\mu, L_\tau$ are not conserved 
(in analogy to the quark sector). 
As a by-product, the $Z$ decay amplitude into two quarks of 
different flavours can be read off from our general expressions.

(ii) The $\nu$SM {\em extended} with {\em one heavy 
ordinary Dirac} neutrino (usual SU(2)$_L\otimes$U(1)$_Y$ quantum numbers). 
This case implies the existence of 
a heavy charged lepton as well. It is a simple application of case 
(i) for heavier neutrinos. Again, total lepton
number $L$ is conserved.

(iii) The $\nu$SM {\em extended} with $n_R=2$ {\em heavy right-handed 
singlet Majorana} neutrinos. 
Not only individual, but also total $L$ is, in general, not conserved
since the presence of Majorana mass terms involves mixing of neutrinos and 
antineutrinos, with opposite lepton number. For two equal and heavy masses 
this case reduces to the addition of {\em one heavy singlet Dirac} 
neutrino. In this latter case $L$ is recovered \cite{Wyler:1983dd}. 

Our results are independent of neutrino mass models.
We take into account constraints on neutrino masses and mixings 
given by oscillation experiments (light sector) or imposed by unitarity and 
precision tests of flavour diagonal and nondiagonal processes
(heavy sector).

\section{$Z\to\ell_1\ell_2$ with massive neutrinos}

The amplitude for the decay of a $Z$ boson into two charged leptons with 
different flavour ($\ell_1$ and $\ell_2$) vanishes in Born approximation
but receives quantum corrections.
It can be written for massless external fermions as:
\bea
{\cal M}
\hspace{-1pt}
=
\hspace{-1pt}
-\frac{ig\alpha_W}{16\pi c_W}{\cal V} 
\varepsilon^\mu_Z \bar u_{\ell_2}(p_2) \gamma_\mu(1
\hspace{-1.3pt}-\hspace{-1.3pt}\gamma_5) u_{\ell_1}(-p_1)
\label{M}
\eea
with $\alpha_W=\alpha/s^2_W$.  The dimensionless form factor ${\cal V}$ is a 
function of $\lambda_Q\equiv Q^2/M^2_W$, with $Q^2=(p_2-p_1)^2$ (to be fixed 
at the $Z$ peak), and the masses $\lambda_i\equiv m^2_i/M^2_W$ and mixings of 
the massive, virtual neutrinos $i$ \cite{Mann:1984dvt,quarkcase}. 
${\cal V}$ gets 
contributions from one-loop vertex- and self-energy-graphs, that are calculated 
here in the 't Hooft-Feynman gauge. The former consist of triangle graphs: two 
virtual neutrinos coupled to the external $Z$ boson with a $W$ or a Goldstone 
boson $\phi$ being exchanged ($W$, $\phi$); and one virtual neutrino exchanged
with $W$ or $\phi$ coupled to the $Z$ ($WW$, $\phi\phi$ and $W\phi$). The
self-energy graphs are corrections to the external fermion legs ($\Sigma$). 

One must discuss separately the cases of {\em ordinary} Dirac and {\em general}
Majorana neutrinos \cite{Illana:1999ww}:

Dirac:
\bea
{\cal V}_D&=& \sum^{n_G}_{i=1} 
{\bf V}_{\ell_1 i} {\bf V}^{^*}_{\ell_2 i}
V(i),
\\
V(i) &=&
v_W(i)+v_{\phi}(i)+v_{WW}(i)+v_{\phi\phi}(i)
\nn\\
&& +v_{W\phi}(i)+v_{\Sigma}(i),
\label{V}
\eea

Majorana:
\bea
{\cal V}_M&=&\sum^{n_G+n_R}_{i,j=1}
{\bf B}_{\ell_1 i} {\bf B}^*_{\ell_2 j}
V_M(i,j),
\\
V_M(i,j)&=&v_W(i,j)+v_{\phi}(i,j)+v_{WW}(i)
\nn\\
&& +v_{\phi\phi}(i)+v_{W\phi}(i)+v_{\Sigma}(i),
\label{FM}
\eea
where only global mixing factors are extracted: the {\em leptonic CKM}
mixing matrix ${\bf V}$ for ordinary Dirac neutrinos, and its {\em generalized} 
version, ${\bf B}$, for Majorana neutrinos. The latter mixings appear in 
charged-current interactions of the left-handed (LH) components of $n_G$ 
generations of ordinary charged leptons ($\ell^0_L = e_L,\mu_L,\tau_L,\dots$) 
with $n_G$ LH isodoublet neutrinos ($\nu^0_L = \nu_e,\nu_\mu,\nu_\tau,
\dots$). The interaction eigenstates are in general not 
the same as the physical mass eigenstates but a mixture \cite{Schechter:1980gr,%
Pilaftsis:1992ug,Ilakovac:1995kj}, 
\bea
{\ell^0_L}_i=\dis\sum_{j=1}^{n_G} {\bf U}^{\ell_L}_{ij}\ {\ell_L}_j, \quad
{\nu^0_L}_i =\dis\sum_{j=1}^{n_G+n_R} {\bf U}_{ij}\ {\nu_L}_j,
\eea
where $\nu$=$\eta\ \nu^c$ are $n_G+n_R$ Majorana fields (self-conjugate up to 
a phase $\eta$). In the physical basis, 
\bea
-{\cal L}_{CC}&=&\frac{g}{\sqrt{2}}W_\mu \overline{\ell^0_L}_i
\gamma^\mu P_L{\nu^0_L}_i + h.c. \nn\\
&=&\frac{g}{\sqrt{2}}W_\mu\ {\bf B}_{ij}\ \overline{\ell_L}_i
\gamma^\mu P_L{\nu_L}_j + h.c.,
\eea
where $P_{R,L}=\frac{1}{2}(1\pm\gamma_5)$ and
\bea
{\bf B}_{ij}\equiv\dis\sum^{n_G}_{k=1} {\bf U}^{\ell^*_L}_{ki}{\bf U}_{kj}
\eea
is a rectangular $n_G\times(n_G\times n_R)$ matrix. 

The main feature distinguishing Dirac and Majorana cases is the existence
of {\em nondiagonal} $Z\nu_i\nu_j$ vertices (flavour-changing neutral currents),
{\em coupling both left- and right-handed} components of the Majorana neutrinos
to the $Z$ boson, 
\bea
-{\cal L}^Z_{NC}
\hspace{-2pt}
&=&
\hspace{-2pt}
\frac{g}{2c_W}Z_\mu
[\overline{\nu^0_L}_i\gamma^\mu P_L {\nu^0_{L_i}} -
 \overline{\nu^{0c}_L}_i\gamma^\mu P_R {\nu^{0c}_{L_i}} ]\nn\\
\hspace{-2pt}
&=&
\hspace{-2pt}
\frac{g}{2c_W}Z_\mu \overline{\nu_i} 
  [{\bf C}_{ij}P_L-{\bf C}^*_{ij}P_R] \nu_j
\eea
with 
\bea
{\bf C}_{ij}\equiv\dis\sum^{n_G}_{k=1} {\bf U}^*_{ki}{\bf U}_{kj},
\quad (i,j=1,\dots,n_G+n_R)
\eea
a quadratic $(n_G+n_R)^2$ matrix.
Such vertices appear in graphs where a $W$ or a Goldstone boson
$\phi$ is exchanged:
\bea
v_W(i,j) & = & 
-{\bf C}_{ij}\big[ \lambda_Q(\c_{0}+\c_{11}+\c_{12}+\c_{23}) \nn\\ &&
\hspace{1cm}
-2\c_{24}+1\big]
\nn \\
                    &   & +{\bf C}^*_{ij}\sqrt{\lambda_i\lambda_j}\ \c_0,
\\ 
v_{\phi}(i,j)  & = &
-{\bf C}_{ij}\displaystyle\frac{\lambda_i\lambda_j}{2}\c_0 
\\   & & 
+ \hspace{-1pt}
{\bf C}^*_{ij}\displaystyle\frac{\sqrt{\lambda_i\lambda_j}}{2}
 \left[\lambda_Q\c_{23}-2\c_{24}+\displaystyle\frac{1}{2}\right]. \nn
\eea
The (diagonal) contributions of these graphs for Dirac virtual neutrinos
[$v_W(i)$ and $v_\phi(i,j)$] are obtained by the replacements:
\bea
\c_{..}\equiv\c_{..}(\lambda_i,\lambda_j)&\to&\c_{..}(\lambda_i,\lambda_i),\\
{\bf C}_{ij}&\to&  (v_i+a_i)\ \delta_{ij} = \delta_{ij},\quad \\
{\bf C}^*_{ij}&\to& -(v_i-a_i)\ \delta_{ij} = 0.
\eea 
The rest of the contributions are:
\bea
v_{WW}(i)  & = & 2c^2_W\ (2I^{i_L}_3) \big[ \lambda_Q\ 
(\cbar_{11}+\cbar_{12}+\cbar_{23}) \nn\\ &&
\hspace{2cm} -6\ \cbar_{24}+1\big],
\\  
v_{\phi\phi}(i)  & = &-(1-2s^2_W)\ (2I^{i_L}_3)\ \lambda_i\
\cbar_{24},
\\  
v_{W\phi}(i)  & = & -2s^2_W\ (2I^{i_L}_3)\ \lambda_i\ \cbar_0,
\\
v_{\Sigma}(i)  & = & 
\displaystyle\frac{1}{2}(v_i+a_i-4c^2_W a_i)\nn\\
&&\hspace{1.5cm}\times\left[(2+\lambda_i)\b_1 +1 \right].
\eea

We have introduced above dimensionless two- and three-point one-loop 
functions:
\bea
\b_1&\equiv&\b_1(\lambda_i) = B_1(0;m^2_i,M^2_W),\\
\bar{\c}_{..}&\equiv&\bar{\c}_{..}(\lambda_i) \nn\\
&=&M^2_W\ C_{..}(0,Q^2,0;m^2_i,M^2_W,M^2_W),\\
\c_{..}&\equiv& \c_{..}(\lambda_i,\lambda_j) \nn\\
&=&M^2_W\ C_{..}(0,Q^2,0;M^2_W,m^2_i,m^2_j),
\eea
from the usual tensor integrals \cite{integrals}
\bea
B^\mu(p^2;m^2_0,m^2_1)=p^\mu B_1, 
\quad\quad\quad\quad
\\
C^\mu(p^2_1,Q^2,p^2_2;m^2_0,m^2_1,m^2_2)\!
=p^\mu_1 C_{11}\! +\! p^\mu_2 C_{12}, \\
C^{\mu\nu}(p^2_1,Q^2,p^2_1;m^2_0,m^2_1,m^2_2)
\quad\quad\quad\quad\quad\quad\quad\  
\nn\\
=p^\mu_1 p^\nu_1 C_{21} + p^\mu_2 p^\nu_2 C_{22} 
+ (p^\mu_1 p^\nu_2 + p^\mu_2 p^\nu_1) C_{23} \nn\\
+ g^{\mu\nu} M^2_W C_{24}.
\eea
The tensor integrals are numerically evaluated  with the computer
program {\tt LoopTools} \cite{TH}.

The expressions above, in the Dirac case, are also valid for quark
flavour-changing $Z$ decays involving virtual quarks $i$, just by substituting 
their corresponding weak isospin $I^{i_L}_3$, electric charge $e_i$, and vector 
$v_i$ and axial-vector $a_i$ couplings,
\bea
v_i&=&I^{i_L}_3 - 2 e_i s_W^2, \\
a_i&=&I^{i_L}_3.
\eea

It turns out convenient to cast (\ref{FM}) as
\bea
V_M(i,j)&=&\delta_{ij}F(\lambda_i)+{\bf C}_{ij}G(\lambda_i,\lambda_j)\nn\\
&&+{\bf C}^*_{ij} \sqrt{\lambda_i\lambda_j} H(\lambda_i,\lambda_j).
\label{VM}
\eea
The Dirac form factor (\ref{V}) is then 
\bea
V(i)=F(\lambda_i)+G(\lambda_i,\lambda_i).
\label{VD}
\eea 
The amplitude ${\cal M}$ is finite without renormalization, but
the form factors $V$ and $V_M$ are not. The divergences are such 
that they exactly cancel due to unitarity relations among the mixing 
matrix elements ${\bf V}$, ${\bf B}$ and ${\bf C}$ 
\cite{Mann:1984dvt,Ilakovac:1995kj,Illana:1999ww}.

The branching ratio $Z\to \ell_1^{\mp}\ell_2^{\pm}$ reads
\bea
{\rm BR}(Z\to \ell_1^{\mp}\ell_2^{\pm})=
\frac{\alpha_W^3M_Z}{192\pi^2c_W^2\Gamma_Z}
 |{\cal V}(M^2_Z)|^2.
\label{BR}
\eea
The constant in front of (\ref{BR}) is of ${\cal O}(10^{-6})$.

\section{Contribution from light neutrinos: $\nu$SM}

Ignoring the not yet confirmed results of the LSND accelerator 
experiment, all neutrino experiments are compatible with the 
oscillation between two of a total of three neutrino species. 
The signals in atmospheric experiments
\cite{Mann:1999zb} are at the 90\% c.l. compatible with
$\nu_{\mu}-\nu_{\tau}$ oscillations,
\bea
\Delta m^2_{\rm atm} &=& \Delta m^2_{23} \simeq 
(2 \div 8) \times 10^{-3} {\rm eV}^2,\\
\sin^2 2\theta_{\rm atm} &=& \sin^2 2\theta_{23} \simeq 
0.82 \div 1.0.
\eea
Solar experiments \cite{Bahcall:1998jt} indicate $\nu_{e}-\nu_{\mu}$ mixing,
\bea
\Delta m^2_{\odot} &=& \Delta m^2_{12} \simeq 
10^{-10} \div 10^{-5} {\rm eV}^2 ,\\
\sin^2 2\theta_{\odot} &=& \sin^2 2\theta_{12}=\mbox{free} 
\eea
(there are solutions for vacuum and matter oscillations compatible
with a wide range of mixing angles). From reactor searches, there are no 
hints of $\nu_{e}-\nu_{\tau}$ oscillations \cite{CHOOZ}, which implies
\bea
\sin^2 2\theta_{13} = 0.
\eea
Taking this information into the standard parameterization for the mixing
matrix \cite{PDG:1998aa} (oscillation experiments are 
insensitive to Majorana CP-phases) and putting the Dirac CP-phase $\delta=0$, 
since no information on it is yet available, one has
\bea
{\bf V} &\simeq&
\left(\ba{ccc}	c_{12} & s_{12} & 0 \\
-\frac{1}{\sqrt{2}}s_{12} & \frac{1}{\sqrt{2}}c_{12} & \frac{1}{\sqrt{2}} \\
\frac{1}{\sqrt{2}}s_{12} & -\frac{1}{\sqrt{2}}c_{12} & \frac{1}{\sqrt{2}} 
\ea\right).
\eea

Using the unitarity of ${\bf V}$, with $\ell_1\neq\ell_2$,
\bea
{\rm BR}(Z\to \ell_1^{\mp}\ell_2^{\pm})
=\frac{\alpha_W^3M_Z}{192\pi^2c_W^2\Gamma_Z}\hspace{2cm} \nn\\
\quad\times \left|\sum^3_{i=1} {\bf V}_{\ell_1 i} {\bf V}^*_{\ell_2 i}
[V(\lambda_i)-V(0)]\right|^2.
\eea
Performing a low neutrino mass expansion of the tensor integrals
($\lambda_i\ll 1$) one finds \cite{Riemann:1982rq,IR}
\bea
V(\lambda_i)-V(0) &=& a_1 \lambda_i + {\cal O}(\lambda^2_i),\\
a_1 &=& 2.5623 - 2.950\ i.
\eea
Therefore BR$(Z\to \ell_1^{\mp}\ell_2^{\pm})$ goes as $m^4_i$
for low neutrino masses. This behaviour is still a good approximation
not far below the $Z$ mass (see Fig.~1).

Substituting the phenomenological squared mass differences 
$\lambda_{ij}\equiv\Delta m^2_{ij}/M^2_W$ and the mixing angles, one has
\cite{Illana:1999ww}
\bea
{\rm BR}(Z\to e^{\mp}\mu^{\pm}) 
\hspace{-3pt}&\simeq&\hspace{-3pt}
{\rm BR}(Z\to e^{\mp}\tau^{\pm}) \nn\\ 
\hspace{-3pt}&\approx&\hspace{-3pt}
6\times 10^{-6}\times c^2_{12}s^2_{12}\lambda^2_{12} \nn\\
\hspace{-3pt}&\lsim&\hspace{-3pt} 
4\times 10^{-60},\\
{\rm BR}(Z\to \mu^{\mp}\tau^{\pm}) 
\hspace{-3pt}&\simeq&\hspace{-3pt} 
3\times 10^{-6}\times
|s^2_{12}\lambda_{12}-\lambda_{23}|^2 \nn\\
\hspace{-3pt}&\approx&\hspace{-3pt} 
(3\div 30)\times 10^{-55}.
\eea
These rates are extremely small. In fact, the contributions from the 
observed light neutrinos will be neglected in the next section, where
we extend the $\nu$SM to accommodate heavy neutrinos, taking massless
the light ones.

\section{Contribution from heavy neutrinos}

\begin{figure}[htb]
\begin{center}
\epsfig{file=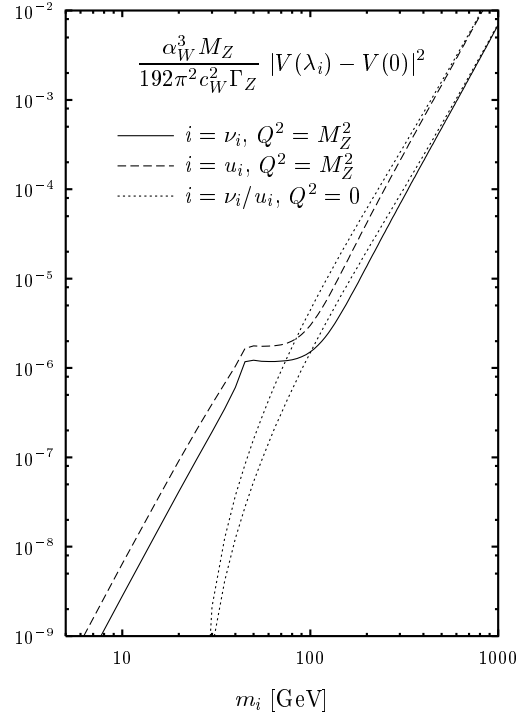,width=0.9\linewidth}
\vspace{-1cm}
\end{center}
\caption{Contribution of one massive Dirac neutrino (solid) or up-quark (dashed)
 to BR normalized to unit mixing. Dotted lines using $Q^2=0$.}
\label{fig: dirac}
\end{figure}

\begin{figure}[htb]
\begin{center}
\epsfig{file=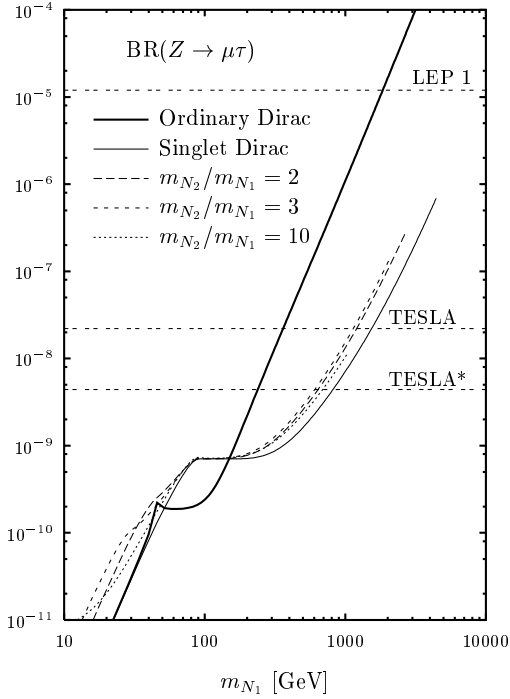,width=0.9\linewidth}
\vspace{-1cm}
\end{center}
\caption{Maximal value of the branching ratio of $Z\to\mu^\mp\tau^\pm$ in 
the $\nu$SM extended with: (i) one heavy ordinary (thick solid) or singlet (thin solid) Dirac 
neutrino of mass $m_{N_1}$; (ii) two heavy right-handed singlet Majorana 
neutrinos (dashed lines) with masses $m_{N_1}$ and $m_{N_2}$.
The upper limits of (\ref{lhmix}) are taken as light-heavy mixings.}
\label{fig: majorana}
\end{figure}

Assume that there is a sector of heavy neutrinos $N_i$ mixing with the light 
ones. Using a general formalism developed in \cite{Langacker:1988ur} one can
exploit measurements of flavour diagonal processes (checks of lepton 
universality and CKM unitarity, $Z$ boson invisible width, ...) 
\cite{nardi} to obtain {\em indirect} experimental 
bounds on light-heavy mixings \cite{Ilakovac:1995kj}:
\bea
s^2_{\nu_\ell} \equiv |\sum_{i} {\bf B}_{\ell N_i}|^2
\label{lhdef}
\eea
(replacing ${\bf B}$ by ${\bf V}$ for Dirac neutrinos).
The most recent indirect bounds \cite{Bergmann:1998rg}:
\bea
s^2_{\nu_e}<0.012,\ 
s^2_{\nu_\mu}<0.0096,\ 
s^2_{\nu_\tau}<0.016,
\label{lhmix}
\eea
are only improved by {\em direct} searches for flavour nondiagonal processes 
involving the first two lepton generations. In fact, from 
BR$(\mu\to e\gamma)<1.2\times 10^{-11}$ \cite{Brooks:1999pu} one may infer 
\cite{Tommasini:1995ii}:
\bea
s^2_{\nu_e}s^2_{\nu_\mu}<1.4\times 10^{-8}.
\eea

\subsection{$\nu$SM + one heavy ordinary Dirac}
From unitarity of ${\bf V}$ for $n_G+1$ generations,
\bea
{\rm BR}(Z\to \ell_1^{\mp}\ell_2^{\pm})&=& 
\frac{\alpha_W^3M_Z}{192\pi^2c_W^2\Gamma_Z}
\left| {\bf V}_{\ell_1N}{\bf V}_{\ell_2N}^*\right|^2\nn\\
&&\times
\left|V(\lambda_N) - V(0)\right|^2.
\eea
The form factor $V$, subtracted, squared and normalized is depicted in 
Fig.~1. The results agree with earlier calculations \cite{Mann:1984dvt}, 
also for the quark flavour-changing $Z$ decays \cite{quarkcase}.
As expected, the approximation $Q^2=0$ is very bad for $m_N\lsim M_Z$ but it 
makes sense for $m_N\gg M_Z$ (Fig.~1). 

Taking the present upper bounds of the mixing matrix elements from
(\ref{lhdef},\ref{lhmix}), Fig.~2 shows the maximal 
BR($Z\to\mu^\mp\tau^\pm$), for illustration.

It is worth noticing that the expansion of tensor integrals in the large 
mass limit \cite{Mann:1984dvt,Illana:1999ww} at the $Z$ peak yields
\bea
V(\lambda_N)-V(0)=\frac{1}{2}[\lambda_N + 2.88 \ln\lambda_N\hspace{1.2cm} \nn\\
\hspace{1.5cm}-(6.99+2.11\ i)] +{\cal O}(\ln\lambda_N/\lambda_N),
\label{dirlim}
\eea
leading again to an $m^4_N$ growth of the branching ratios for large
neutrino masses.

\subsection{$\nu$SM + $(n_R=2)$ Majorana neutrinos}

Unitarity constraints on ${\bf B}$ and ${\bf C}$ \cite{Ilakovac:1995kj} 
allow to write ${\cal V}_M$ in terms of the heavy sector only:
\bea
{\cal V}_M&=&\sum^{n_R}_{i,j=1}{\bf B}_{\ell_1 N_i} {\bf B}^*_{\ell_2 N_j}
\nn\\
&\times&\Big\{\delta_{N_i N_j}
[\ F(\lambda_{N_i})-F(0)+G(\lambda_{N_i},0)
\nonumber\\ &&\quad\quad\quad\quad +G(0,\lambda_{N_i})-2G(0,0)]  \nonumber\\
&&+{\bf C}_{N_i N_j}
[\ G(\lambda_{N_i},\lambda_{N_j})-G(\lambda_{N_i},0)\nonumber\\
&&\quad\quad\quad\quad-G(0,\lambda_{N_j})+G(0,0)] \nonumber\\
&&+{\bf C}^*_{N_i N_j} \sqrt{\lambda_{N_i}\lambda_{N_j}} \
H(\lambda_{N_i},\lambda_{N_j})\Big\}.
\eea

For $n_R=2$ the mixing matrices are exactly calculable in terms of 
$s^2_{\nu_\ell}$ and $r\equiv m^2_{N_2}/m^2_{N_1}$ \cite{Ilakovac:1995kj}.
The upper values for the branching ratios can be then straightforwardly 
obtained from the bounds (\ref{lhmix}), given the heavy masses $m_{N_1}$, 
$m_{N_2}$ (Fig.~2). The case $m_{N_1}=m_{N_2}$ is equivalent to {\em one 
heavy singlet Dirac} neutrino (in fact, two equal mass Majorana neutrinos
with opposite CP parities form a Dirac neutrino).

In the large neutrino mass limit ($\lambda_{N_1}\gg 1$) one obtains
\cite{IR}
\bea
{\cal V}_M(\lambda_{N_1},r;s_{\nu_\ell})
=s_{\nu_{\ell_1}} s_{\nu_{\ell_2}} \hspace{2.8cm}
\nn\\
\times\Bigg\{\frac{\sum_\ell s^2_{\nu_{\ell}}}{(1+r^\frac{1}{2})^2}
\left(\frac{3}{2}r+\frac{r^2+r-4r^\frac{3}{2}}{4(1-r)}
\ln r\right)\lambda_{N_1} 
\nn\\
+\frac{1}{2}\left(3+\frac{1-2c^2_W}{6}\lambda_Q\right)
\ln\lambda_{N_1}\Bigg\} + {\cal O}(1).
\label{majlim}
\eea
The constant in front of the $\ln\lambda$ term is identical to the one
in the Dirac case (\ref{dirlim}).
In the approximation $\lambda_Q\equiv Q^2/M^2_W=0$, 
the expression (\ref{majlim}) is in agreement with \cite{Ilakovac:1995kj},
but we take the actual value $\lambda_Q=M^2_Z/M^2_W=1/c^2_W\approx 1.286$ 
in the whole calculation.

Finally, notice that large neutrino masses are restricted by the perturbative 
unitarity condition on the decay width of heavy neutrinos \cite{Fajfer:1998px},
\bea
\Gamma_{N_i}\simeq(2\times)
\frac{\alpha_W}{8M^2_W}m^3_{N_i}\sum_{\ell}|{\bf B}_{\ell N_i}|^2\leq 
\frac{1}{2} m_{N_i}
\label{pub}
\eea
for Dirac (Majorana) neutrinos. In this way,
the unacceptable large-mass behaviour of the amplitudes ($\propto m^2_N$)
is actually cured when a sensible light-heavy mixing (at most 
$\propto m^{-2}_N$) is taken into account \cite{decoupling}. The restrictions
(\ref{pub}) lead to the end-points in the curves of Fig.~2.

\section{Summary and conclusions}

The perturbative unitarity limit on the decay width of heavy neutrinos
effectively prevents their nondecoupling, thus ensuring
smaller light-heavy mixings for increasing heavy masses.
Given the present {\em indirect} upper bounds on light-heavy mixings 
of ${\cal O}(10^{-2})$,
this already sets indirect upper limits on LFV $Z$ decays 
of about BR$\lsim 10^{-6}$ with Majorana neutrinos
and above LEP 1 reach for Dirac neutrinos.

We have presented the full one-loop expectations for the {\em direct} 
lepton flavour changing
process $Z\to\ell_1\ell_2$ with virtual Dirac or Majorana neutrinos.
We conclude that:
(i) the contributions from the observed light neutrino sector are
far from experimental verification (BR$\lsim 10^{-54}$);
(ii)  the Giga$-Z$ mode of the future Tesla linear
collider, sensitive down to  about BR$\sim 10^{-8}$, 
might have a chance to produce such processes.

\end{document}